\documentclass{article}

\usepackage{arxiv}

\usepackage[utf8]{inputenc} % allow utf-8 input
\usepackage[T1]{fontenc}    % use 8-bit T1 fonts
\usepackage{hyperref}       % hyperlinks
\usepackage{url}            % simple URL typesetting
\usepackage{booktabs}       % professional-quality tables
\usepackage{amsfonts}       % blackboard math symbols
\usepackage{nicefrac}       % compact symbols for 1/2, etc.
\usepackage{microtype}      % microtypography
\usepackage{lipsum}
\usepackage{graphicx}
\graphicspath{ {./images/} }
\usepackage{comment}

\title{Undulating patterns of Hysteresis loops in diurnal seasonality of air temperature in Urban Heat Island effect: Insights from Paris and Madrid}

\author{
 Suman Dharmasthala \\
  Independent Researcher\\
  \texttt{suman.nad@gmail.com  } \\
   \And
 Vittal Hari \\
  IIT (ISM) Dhanbad\\
  \texttt{vittal@iitism.ac.in} \\
  \And
 Rohini Kumar \\
  Helmholtz Centre for \\ Environmental Research - UFZ\\
  \texttt{rohini.kumar@ufz.de} \\
}

\begin{document}
\maketitle
\begin{abstract}
This study examines the dynamics of the urban heat island (UHI) effect by conducting a comparative analysis of air temperature hysteresis patterns in Paris and Madrid, two major European cities with distinct climatic and urban characteristics. Utilizing high-resolution modelled air temperature data aggregated at a fine temporal resolution of three-hour intervals from 2008 to 2017, we investigate how diurnal and seasonal hysteresis loops reveal both unique and universal aspects of UHI variability. Paris, located in a temperate oceanic climate, and Madrid, situated in a cold semi-arid zone, display pronounced differences in UHI intensity, seasonal distribution, and diurnal patterns. Despite these contrasts, both cities exhibit remarkably similar hysteresis loop directions and slopes, suggesting that time-dependent mechanisms such as solar radiation and heat storage fundamentally govern air temperature UHI across diverse urban contexts. Our findings underscore the importance of considering both local climate and universal physical processes in developing targeted, climate-resilient urban strategies. The results pave the way for group-based interventions and classification of cities by hysteresis patterns to inform urban planning and heat mitigation efforts. 
\end{abstract}

% keywords can be removed

\section{Introduction}
The urban heat island effect (UHI) refers to the phenomenon in which urban areas experience higher temperatures than their surrounding rural or background (non-urban) regions, resulting in island-like isotherms in air and surface temperature patterns \cite{oke2017urban}. This effect is a well-documented consequence of urbanisation, driven by increased built surfaces, reduced vegetation, and anthropogenic heat emissions \cite{arnfield2003two, oke1982energetic}. The effect of UHI is not just a climatic anomaly \cite{pappalardo2023mapping}; It poses significant risks to human health, including increased morbidity and mortality related to heat \cite{Giannopoulou2014Feb, Santamouris2020Jan}, and exacerbates energy consumption and environmental degradation \cite{Ihara2008Jan, Santamouris2020Jan, yang2020impact}. Given its detrimental impacts \cite{kim2021urban}, extensive research has been devoted to understanding the effect of UHI to develop effective mitigation and adaptation strategies at urban scales \cite{Deilami2018May, Ulpiani2021Jan}. 

Despite the wealth of studies on UHI, a universal mitigation strategy remains obscure \cite{manoli2019magnitude}, generally due to its dependence on a multitude of factors, including local climatic zones, latitude, elevation, city size, land cover, and urban morphology \cite{tzavali2015urban, schwarz2015analyzing, Deilami2018May}. For instance, the magnitude and pattern of UHI can vary significantly between cities in temperate and tropical climates \cite{manoli2019magnitude}, underscoring the need for comparative studies. Furthermore, while numerous studies have proposed mitigation measures at the local or individual city level, there is a lack of consensus on scalable and transferable solutions \cite{manoli2019magnitude}. This gap highlights the importance of understanding the underlying factors driving UHI variability across different urban contexts. 

Most studies investigating the UHI effect focus on temperature differences between urban and non-urban (rural or background) areas, although the definitions of 'urban' and 'non-urban' vary widely across the literature \cite{schwarz2011exploring}. These studies typically employ either surface temperature measurements, derived from remote sensing, or air temperature measurements, obtained from weather stations \cite{oke2017urban, yang2020comparison}. Each approach has its limitations: surface temperature studies often analyse distinct time periods (day and night) but lack temporal resolution, while air temperature studies, though temporally rich, are spatially sparse and often rely on short-term observational data \cite{kim2021urban, kim2021urban_1}. In cases where finer temporal resolutions are considered, hysteresis analysis - a critical tool for understanding the lagged response of UHI to ambient temperature changes - is rarely performed \cite{memon2009investigation, hu2019assessment, shi2018modelling, cleland2023urban, li2018impact}. This limitation stems from the trade-off between spatial and temporal resolution in UHI data: remotely sensed surface temperature data offers high spatial resolution but limited temporal coverage, whereas air temperature data from weather stations provides continuous temporal coverage, but with poor spatial coverage \cite{sficua2023surface}. 

Studying the UHI effect at finer spatial and temporal resolutions, though computationally demanding, can reveal one of its most intriguing signatures: seasonal hysteresis loops. These loops, generated by plotting the case of surface UHI against rural or background temperature, exhibit rate-dependent seasonality that is closely tied to local climatic conditions \cite{sismanidis2022seasonality, manoli2020seasonal}. Interestingly, while surface UHI hysteresis is influenced by factors such as land cover and urban morphology, air temperature hysteresis appears to be primarily a function of time of the day \cite{zhou2016assessing}. Despite this distinction, research on hysteresis loops has predominantly focused on surface UHI, with limited attention given to air temperature hysteresis \cite{zhou2016assessing, song2017hysteresis, zhou2013statistics}. This gap in the literature is significant, as air temperature hysteresis, studied at a fine temporal resolution, provides critical insights into the diurnal and seasonal dynamics of UHI, which are essential for developing effective mitigation strategies, which this study seeks to address. 

Comparative studies of UHI dynamics across cities with differing climatic and urban contexts have provided valuable insights into factors driving UHI variability. For example, \cite{sismanidis2022seasonality} examined the hysteresis pattern in cities across Koppen climate classification \cite{Beck2018}, highlighting how variations in local climatic conditions influence UHI hysteresis patterns for surface temperature. Such studies underscore the importance of context-specific approaches to UHI mitigation, as strategies effective in one climatic zone may not translate well to another. In this study, we focus on a comparative study of air temperature hysteresis effect for Paris and Madrid, two cities with well-documented UHI characteristics but distinct climate and urban contexts. Paris, located in a Temperate, CfB Koppen climate classification, exhibits surface UHI patterns looping in an upward direction, while Madrid, situated in an Arid BsK classification [Fig. 1], experiences a similar clockwise, but downward sloping hysteresis loop, attributing this distinction of hysteresis patterns among the two cities to time lags between temperature and rainfall \cite{manoli2020seasonal}. While the study looked at surface temperature UHI (SUHI), we aim to look at fine resolution air temperature UHI (termed as UHI throughout the paper) as a comparative study to enhance our understanding of the hysteresis pattern, which is limited in air temperature data. 

To address these limitations, this study leverages modelled fine temporal resolution hourly air temperature data (aggregated at three-hour intervals for clarity while preserving the patterns) for the two cities: Paris and Madrid. These cities were selected due to their UHI research history, enabling meaningful comparisons with existing studies. By focusing on two cities, we also mitigate the computational challenges associated with acquiring and processing high-resolution temporal data while providing a detailed analysis of diurnal and seasonal hysteresis patterns. Our findings contribute to a deeper understanding of UHI dynamics and offer insights into the factors driving hysteresis variability across different urban contexts for air temperature data. 

\section{Data and Methodology}

\subsection{Study Area}
Paris and Madrid, two of Europe's most prominent metropolitan areas, were selected for this study because of their contrasting climatic and urban characteristics, which provide a robust framework for investigating Urban Heat Island (UHI) dynamics and hysteresis patterns. Paris, the capital of France, is located in the northern part of the country (48.8566° N, 2.3522° E) and is classified under the Temperate Oceanic Climate (Cfb) according to the Köppen-Geiger climate classification. This climate is characterised by mild summers, cool winters, and a relatively even precipitation distribution throughout the year. The urban landscape of the city is marked by dense building configurations \cite{lemonsu2015vulnerability}, moderate vegetation cover \cite{larondelle2013urban}, and a population density of approximately 21,000 inhabitants per square kilometer \cite{insee2020population}.  

\begin{figure} 
    \centering
    \includegraphics[width=\textwidth]{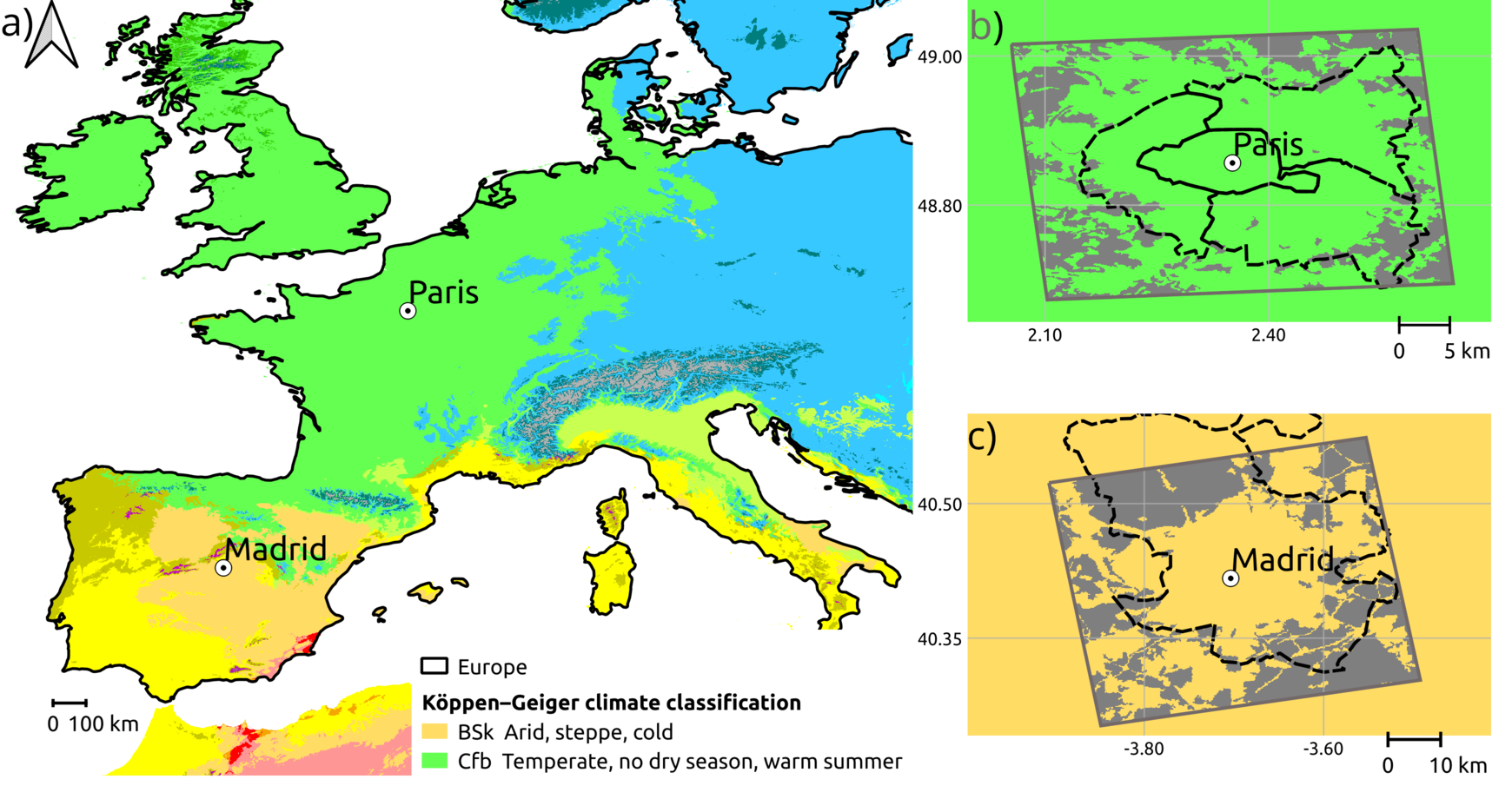}
            \caption{The map on the left (1a) contains the location of Paris and Madrid in Europe; colours show the Koppen-Gieger climate classification. The maps on the right (1b and 1c) contain the cities of Paris (top) and Madrid (bottom) with city boundaries; the urban mask is shown in the colour of the climate classification within the rural mask box represented by grey (area outside the grey bounds is not of significance in the study).}
\end{figure}

In contrast, Madrid, the capital of Spain, is situated in the central part of the Iberian Peninsula  (40.4168° N, 3.7038° W) and falls under the Cold Semi-Arid Climate (BSk) in the Köppen-Geiger classification. This climate is defined by hot, dry summers, cool winters, and limited annual precipitation. Madrid’s urban structure is characterised by expansive built-up areas \cite{sobrino2012impact}, lower vegetation density compared to Paris \cite{gomez2004experimental}, and a population density of around 5,300 inhabitants per square kilometer \cite{ine2020population}.

Fig. 1a shows the location of Paris and Madrid in the current Koppen-Geiger climate classification map of Europe. The urban and rural (background) masks provided in Fig. 1b and 1c, show the extent of urban and background areas considered in this study for both cities. The boundaries of both the Greater Paris and Madrid city are also marked for reference in Fig. 1b and 1c. The area as given by the city boundaries is 814 sq km for Métropole du Grand Paris and 604 sq km for Madrid City (Municipality), while the bounding box boundaries of both are 1600 sq km and 900 sq km, respectively. The contrasting climatic conditions and physical features of these two cities, which play a key role in shaping surface UHI dynamics \cite{manoli2020seasonal, tzavali2015urban, schwarz2015analyzing, Deilami2018May} and its hysteresis pattern in particular, provide a compelling rationale for the comparative study undertaken here for air temperature UHI and its hysteresis pattern.

\subsection{Data}

The study uses modelled air temperature (2m above surface) data from the Copernicus Climate Change Service website \cite{Climatev30:online}, which is available for Paris, Madrid and other European cities. The data has a spatial resolution of 100 mt and a temporal resolution of one hour, spanning from January 2008 to December 2017. This data is modelled using the UrbClim model, which downscales the ERA5 reanalysis data and incorporates comprehensive meteorological parameters such as latent heat flux, wind velocity, precipitation, and terrain parameters such as vegetation index, land use, and anthropogenic heat flux \cite{de2015urbclim}. 

The spatial extent of the data is defined by a bounding box which covers most of the built-up areas of the cities and their surroundings. The bounding box is visually represented by the box made by the extent of the grey areas in Fig. 1b and 1c, where the grey denotes background (rural) regions, and the non-grey areas within the bounding box represent urban areas. These urban-rural masks are also provided by the UrbClim dataset alongside the 2 mt air temperature data.  

%\bigskip

\subsection{Methodology}

To analyse the urban heat island (UHI) effect, we calculate the weighted average of the hourly air temperature data using pixel areas of the rural and urban masks. This process yields two time series of hourly data, one for the urban areas and one for the background (non-urban) areas. Although hourly data is available for both cities, we average the data over three-hour intervals to preserve observed patterns while facilitating easier visualisation. 

The time series data of spatial average for urban and rural (background) masks are used in the study. The UHI is calculated as the difference between the spatial averages of urban temperature and the background (non-urban) temperature, represented by the formula: 

%<Tuhi> = <Tu> - <Tb>
$$ <T_{uhi}> = <T_u> - <T_b> $$

Here, $<T_{uhi}>$ is the weighted average of urban heat island difference, $<T_u>$ is the spatially aggregated urban temperature, and $<T_b>$ is the non-urban temperature at a three hourly timescale. Weights are based on the pixel counts of the urban and rural masks. 

Hysteresis plots are created by averaging the data across the years (2008 to 2017) and all months for every 3 hours and plotting delta T (UHI) against the background temperature. This approach of plotting the hysteresis reveals both seasonal and diurnal patterns. 

Visual comparisons of the hysteresis loops between Paris and Madrid are conducted to highlight the similarities and differences in their UHI dynamics. One limitation of the dataset is that the urban and rural areas are not clearly demarcated, which is a common issue in UHI studies \cite{schwarz2011exploring}. The domain includes areas surrounding the urban zones, which may not be purely rural but may be peri-urban, where air temperature UHI effects can still be observed due to factors such as wind direction.  

\section{Results and Discussion}

Paris and Madrid are located in two distinct climatic zones according to the Koppen climate classification \cite{Beck2018} [Fig. 1], namely Cfb (Temperate Oceanic) and BSk (Cold Semi-Srid), respectively. These climatic differences, combined with their unique urban characteristics, result in pronounced variations in UHI intensity both seasonally and diurnally [Fig. 2a and 2b]. In Paris, the UHI is generally more intense than in Madrid during the day and in seasons others than winter. This is consistent with the city's temperature climate, where moderate humidity and balanced temperature fluctuations contribute to higher UHI intensity \cite{lemonsu2015vulnerability}. In contrast, Madrid's arid climate, characterised by high solar radiation and low humidity, leads to stronger nighttime UHI effects, particularly during summer months \cite{sobrino2012impact}. 

The trimodal distribution of UHI intensity (day, night and others) is observed in both cities, but the patterns differ significantly. While both cities exhibit daytime UHI intensities consistently lower than nighttime values, and greater variability at night compared to the day, their seasonal distributions differ. Specifically, Paris shows reduced winter nighttime variability, whereas Madrid maintains high variability across all seasons [Fig. 2c]. These observations align with prior studies, which attribute daytime UHI primarily to solar radiation, while nighttime UHI is influence by the slow release of stored heat from urban surfaces, the canyon effect of buildings, climatic conditions (particularly in the winter), and evapotranspiration \cite{oke1982energetic, arnfield2003two, stewart2012local} 

Isotherm plots, which visualise both diurnal and seasonal variations in UHI simultaneously [Fig. 2d], reveal further distinctions between the two cities. Madrid exhibits an 'oasis effect' during daytime, where urban temperatures are cooler than its background temperatures, resulting in a negative UHI during daytime across several months (on average). Nighttime UHI also differ markedly between the two cities, reflecting the influence of their respective climates. 

Despite these differences in the UHI characteristics driven by local climate, urban morphology, and seasonal factors, we find that both cities exhibit remarkably similar hysteresis patterns when UHI intensity is analysed against background temperatures. While isotherm plots highlight distinct diurnal and seasonal behaviours - such as the 'oasis effect' and Paris's less variable winter nighttime UHI - the underlying hysteresis loops reveal similar directional patterns (clockwise/anticlockwise) and slopes. The changes seen are in terms of shapes of the loops (and thickness). This suggests that while local features (eg, Madrid's arid climate or Paris's dense urban morphology) modulate the air temperature UHI magnitude and variability, the fundamental processes governing air temperature UHI hysteresis - follow universal principles across climatic contexts.    

This intriguing pattern of formation of hysteresis loops when UHI intensity is plotted against background temperature is observed from previous UHI studies (mostly undertaken for surface temperature UHI hysteresis) \cite{zhou2013statistics, zhou2016assessing, song2017hysteresis, manoli2020seasonal, sismanidis2022seasonality, yang2023diverse, sismanidiswork}. Hysteresis loops provide a comprehensive visualisation of seasonal and diurnal UHI patterns simultaneously, offering a more intuitive understanding than isotherm plots or other uni-variable plots. In this study, we observe that while the mean hysteresis loops for Paris and Madrid are distinct only with respect to their shapes and thickness [Fig. 3e and 3f], they share similarities in the direction (clockwise/anti-clockwise) and slope of the loops [Fig. 3a, 3b, 3c and 3d]. These similarities suggest that certain aspects of air temperature UHI hysteresis are governed by time-dependent mechanisms, such as solar radiation and heat storage and emissivity, while others are influenced by local climatic and urban factors \cite{zhou2016assessing}.  

\begin{figure} 
    \centering
    \includegraphics[width=\textwidth]{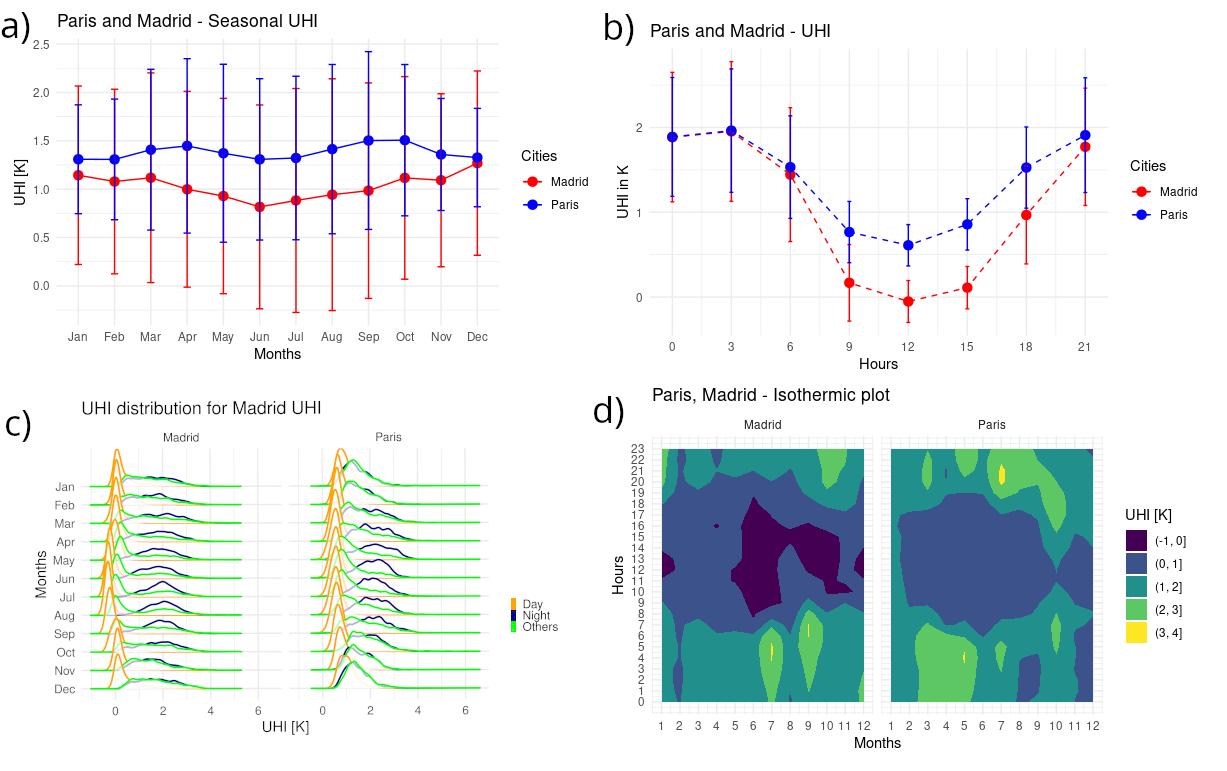}
            \caption{(a) shows similar seasonal peaks and troughs in cities of Paris and Madrid UHI, although Paris is much higher and Madrid has more variability and also has 'Oasis effect' (urban cool island effect). (b) shows similar results for diurnal data, but UHI variability is high in both cities at night. (c) shows the trimodal peaks for the different time groupings [daytime is taken as 09:00 to 17:00 hrs, night time is taken as 21:00 to 05:00 hrs, and others taken as 18:00 to 20:00 hrs (dusk) and 06:00 to 08:00 (dawn)]. (d) shows that seasonal and diurnal patterns are different for both cities.}
\end{figure}

Zhou et. al. \cite{zhou2016assessing}, in their study of London argue that hysteresis effects are largely absent in air temperature UHI due to its lower magnitude (0.5°C to 2°C) compared to surface UHI (0.5°C to 4.5°C). They also suggest that air temperature hysteresis is primarily a function of the time of the day rather than local climatic or physical features. While our findings confirm that the magnitude of air temperature UHI is indeed smaller (approximately 2.5°C), we also observe distinct loop shapes (thickness and length) for Paris and Madrid. This indicates that while the time of the day plays a significant role, local factors such as urban morphology and climate cannot be entirely discounted. It would be interesting to conduct future research to explore the individual shapes of hysteresis loops across multiple cities and compare them with respective surface UHI at fine temporal resolution for both air and surface temperature UHI hysteresis. This would provide a deeper insight into these dynamics and help us with our understanding of UHI and subsequently devise mitigative policies. 

The observed hysteresis loop patterns for both cities reveal several key insights [Fig. 3]. First, nighttime loops are upward-sloping, indicating that winter nighttime UHI is significantly lower than summer nighttime UHI. This is consistent with the greater heat storage and slower release in urban areas during the summer nights \cite{oke1982energetic}. Nighttime is this study is defined as 21:00 to 05:00 hours. Second, daytime loops are downward-sloping, reflecting higher UHI intensity during winter days compared to summer days. This pattern in attributed to reduced solar radiation and lower background temperatures in winter, which amplify the relative UHI effect. Daytime is defined as 09:00 to 17:00 hours. 

\begin{figure} 
    \centering
    \includegraphics[width=\textwidth]{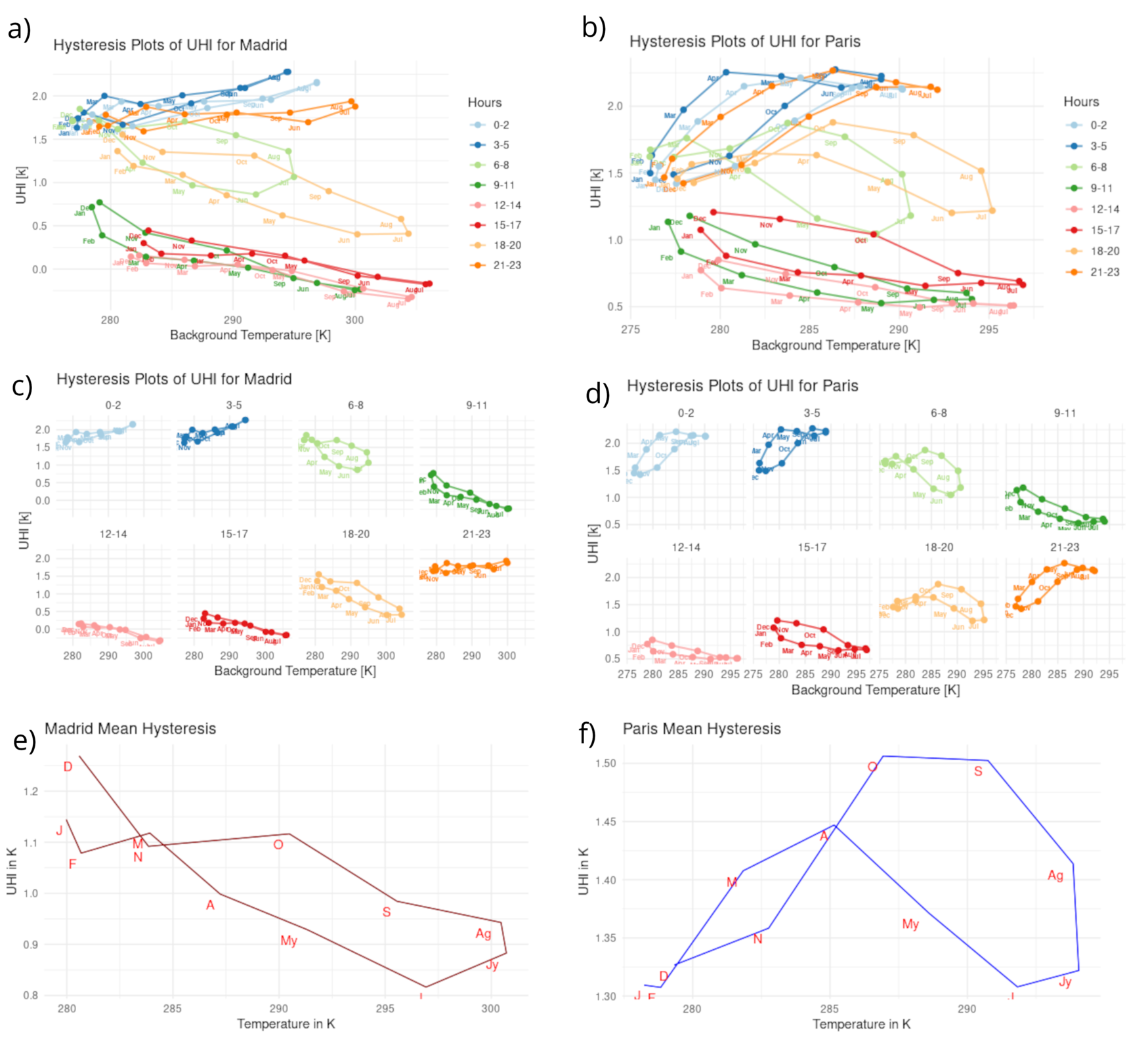}
            \caption{(a) and (b) show hysteresis plots of seasonal diurnality (3 hourly); (c) and (d) show the individual time-period (averaged over 3 hours) plots for easy visualisation, and (e) and (f) show the mean hysteresis averaged across the day for Paris and Madrid respectively.}
\end{figure}

Third, nighttime loops exhibit a clockwise direction, indicating that spring UHI is more intense than fall UHI for the same background temperature. This suggests a lag in the release of stored heat from urban materials during spring. Fourth, daytime loops exhibit an anti-clockwise direction, meaning fall UHI is more intense than spring UHI for the same background temperature. This pattern may be driven by differences in solar radiation and vegetation activity between the two seasons. These findings align with the work of Zhou et al. \cite{zhou2013statistics}, who first identified the unique hysteresis signatures of cities based on seasonal and diurnal UHI variations. 

Finally, dawn (06:00 to 08:00 hours) and dusk (18:00 to 20:00 hours) loops display a twisted '8' shape, similar to the hysteresis patterns observed in surface-air temperature studies over concrete pavements during winter \cite{song2017hysteresis}. This phenomenon is likely due to the transitional nature of these periods, where rapid changes in solar radiation and heat flux occur. The general patterns and directions of hysteresis loops Madrid are flatter compare to Paris, but they follow similar slopes and movements (clockwise and anti-clockwise). This reinforces the idea that air UHI hysteresis is primarily a function of the time of day. However, the differences in loop thickness and length of the loops between the two cities suggest that local climatic and urban features also play a role. As noted by Oke \cite{oke1982energetic}, while surface UHI depends heavily on local features such as land cover and urban morphology, air UHI is more influenced by radiative processes and time lags, particularly during nighttime when heat storage effects are most pronounced. 

The observed hysteresis patterns can be attributed to several underlying mechanisms. During the day, solar radiation heats urban surfaces, which then transfer heat to the surrounding air, creating a strong UHI effect. At night, the release of stored heat from urban materials maintains higher air temperatures, leading to nighttime UHI. In Paris, the temperate climate and moderate humidity levels result in a more balanced heat exchange, with pronounced daytime UHI and stable nighttime UHI. In contrast, Madrid's arid climate and low humidity enhance radiative cooling at night, leading to stronger nighttime UHI. These mechanisms are further influenced by urban geometry, with dense building configurations in Paris promoting heat retention, while Madrid's expansive built-up areas facilitate heat dissipation during the day but exacerbate heat retention at night \textbackslash{}cite\{oke1982energetic\}.

The findings of this study have important implications for urban planning and UHI mitigation. Given the slightly different shapes of the hysteresis loops, local climate-specific interventions may be required. In Paris, strategies such as increasing vegetation cover and implementing reflective surfaces may help moderate daytime UHI intensity and reduce heat stress during summer months \cite{larondelle2013urban}. In Madrid, more aggressive measures, such as urban greening and water-based cooling systems, may be necessary to address the intense nighttime UHI effects driven by its arid climate \cite{gomez2004experimental}. These context-specific approaches can enhance urban climate resilience and improve the livability of cities in different climatic zones. However, given the significant time-dependent nature of the hysteresis patterns and similarity of loops between the cities, diurnal and seasonal interventions for both cities may be required. 

While this study focuses on Paris and Madrid, the findings can be contextualised by comparing them with UHI patterns in other cities. For example, cities in tropical climates, such as Singapore, may give us an even better picture of the hysteresis loop patterns based on the differences and similarities. In addition to the geographic limitation of the study (study of Europe), this study has other limitations. The reliance on hourly air temperature data from modelled data, may differ from observations from weather stations, although modelled data provides for a high spatial resolution as well as fine temporal resolution. Another limitation of this study is that urban and rural are not clearly demarcated, which is a limitation in most studies of UHI \cite{schwarz2011exploring} and the domain includes the area surrounding the urban and may not necessarily be ‘rural’ in the proper sense (i.e., may be peri-urban, where the UHI effect may still be seen due to wind direction, etc.).

\section{Conclusions}

This study demonstrates that UHI dynamics in Paris and Madrid are governed by both local and universal time-dependent processes. While there is no one-size fits all UHI mitigation strategy, our study shows that a fine temporal resolution hysteresis study allows us to better understand the undulating patterns of air UHI as a function of time, which shows that certain aspects of air temperature UHI are indeed similar. 

Local climatic and urban features—such as Madrid’s arid climate amplifying nighttime UHI variability and Paris’s temperate stability—drive distinct diurnal and seasonal patterns visible in isotherm plots. However, the similarities observed in hysteresis signatures (e.g., clockwise nighttime loops, anti-clockwise daytime loops) point to invariant mechanisms like solar radiation and heat storage that transcend local climatic contexts.

This paves the way for further research into fine-temporal resolution air as well as surface temperature UHIs, to offer new insights for targeted intervention strategies or even for classification of cities and thus their mitigation strategies based on the hysteresis patterns for grouped targeted interventions. By reconciling local and universal UHI drivers, this work provides a framework for climate-adaptive cities, urging planners to pair localised solutions (e.g., Madrid’s water-sensitive design) with global best practices (e.g., heat-retentive materials).

\bigskip
\bigskip
\bigskip
\bigskip

\subsection*{Acknowledgment}

We thank Prof. Martin Hanel, Czech University of Life Sciences Prague, for their initial review of this work.

\clearpage

\bibliographystyle{unsrt}  
\bibliography{references}

\end{document}